\documentclass[12pt]{iopart}
\usepackage{graphicx}
\begin{document}

\def\sNN{$\sqrt{s_{_{NN}}}$}

\title[Lumpy Initial Conditions]{Implications of space-momentum correlations and geometric fluctuations in heavy-ion collisions}

\author{Paul Sorensen}

\address{Brookhaven National Laboratory, P.O. Box 5000, Upton, NY 11973}
\ead{prsorensen@bnl.gov}

\begin{abstract}
  The standard picture of heavy-ion collisions includes a collective
  expansion. If the initial energy density in the collisions is lumpy,
  then the expansion can convert the spatial lumpiness into
  correlations between final-state particles. Correlations
  in heavy-ion collisions show prominent features
  not present in p+p collisions. I argue that many features
  of these correlations are related to the transference of
  over-densities from the initial overlap region into momentum-space
  during the QGP phase of the expansion. I show results
  from a toy Monte-Carlo to illustrate the consequences of lumpy
  initial conditions and a collective expansion.
\end{abstract}


\submitto{\JPG} 

\section{Introduction}\label{intro}
Heavy ionized nuclei are collided at high energies in laboratory
experiments to recreate temperatures and densities similar to those in
the early universe at approximately one microsecond after the big
bang~\cite{wp}. At that time, the universe was filled with deconfined
quarks and gluons, a phase of matter known as the quark-gluon
plasma. Studying the quark-gluon plasma in the laboratory
involves observing hadrons emitted from the fireball created in the
collision of ultrarelitivistic heavy nuclei. The fireball is
$10^{-14}$ meters across and expands for approximately $5 \times
10^{-23}$ seconds before hadrons form, re-scattering ceases, and the
stable particles stream away from the collision zone to eventually be
detected by particle detectors many of orders of magnitude larger than
the speck of quark-gluon-plasma created in the
collisions. Understanding the evolution of the matter created
in those collisions by observing the hadrons that stream into the
detectors is a challenge.

In this talk I discuss how fluctuations in the geometry of the initial
overlap zone~\cite{ic} can manifest as two-particle
correlations. These correlations may therefore provide information
about the expansion phase that converts the over-densities in the
initial overlap zone into momentum space correlations in the
final-state. I use a toy Monte-Carlo model to illustrate the relevance
of $v_n$ fluctuations where $v_n$ are the coefficients in a Fourier
expansion of the particle multiplicity with respect to the reaction
plane. It's important to consider the odd values of $n$ particularly
since some of the interesting features of recent two particle
correlations data from RHIC can be readily understood in terms of
$v_3$ fluctuations. An analysis of $\sqrt{\langle v_n^2 \rangle}$ to
study the expansion of heavy-ion collisions was first proposed in
Ref.~\cite{Mishra:2007tw}. $v_2$ fluctuations and di-hadron
correlations are discussed in Ref.~\cite{Trainor:2007ny} where $v_n$
fluctuations for $n>2$ are argued to be zero and the remaining
structures in correlations data are argued to arise from jets.

\section{The Correlation Landscape at RHIC}

Correlations and fluctuations have long been considered a good
signature for Quark Gluon Plasma (QGP) formation in heavy-ion
collisions~\cite{fluctuations}. Early proposals for QGP searches
suggested searching for a non-monotonic dependence of fluctuations on
variables related to the energy systems density ({\it e.g.}
center-of-mass energy or collision centrality) the expectation being
that above some energy density threshold, a phase transition to QGP
would occur. Data from the experiments at RHIC indeed reveal
interesting features in the two-particle correlation landscape
\cite{onset,ridgedata} but that data does not seem to support a
picture based on correlations arising from a phase
transition. Correlation structures unique to Nucleus-Nucleus
collisions are found but their longitudinal width demonstrates that
they come from the earliest moments of the
collisions~\cite{DGMV}. While two-particle correlations in p+p and
d+Au collisions show a jet-peak narrow in relative azimuth
$\Delta\phi$ and pseudo-rapidity $\Delta\eta$, the near-side peak in
Au+Au collisions broadens substantially in the longitudinal direction
and narrows in azimuth~\cite{onset}.

An analysis of the width of the peak for particles of all $p_T$ finds
the correlation extends across nearly 2 units of pseudo-rapidity
$\eta$~\cite{onset,ridgedata}. When triggering on higher momentum particles
($p_T>2$~GeV/c for example), the correlation extends beyond the
acceptance of the STAR detector ($\Delta\eta<2$) and perhaps as far as
$\Delta\eta=4$ as indicated by PHOBOS data~\cite{ridgedata}.  It has
been proposed that the ridge-like correlations are related to either
non-perturbative multi-quark or gluon effects on mini-jets in Au+Au
collisions~\cite{minij}, to soft gluons radiated by hard partons
traversing the overlap region~\cite{Maj}, to initial spatial
correlations in the system converted to momentum-space correlations by
a radial Hubble expansion~\cite{radflow}, to beam-jets also boosted by
the radial expansion~\cite{beamjets}, or to viscous
broadening~\cite{visc}.

The modifications to the correlations are not limited to small
$\Delta\phi$. When at least one of the particles used in the analysis
has $p_T>2$~GeV/c (a selection made to attempt to increase sensitivity
to jets), the correlation structure at $\Delta\phi>\pi/2$ (away-side)
is also highly modified in comparison to $p+p$
collisions~\cite{awayside}. Instead of a narrow peak with a maximum at
$\Delta\phi=\pi$ corresponding to an away-side jet, a peak shifted
away from $\Delta\phi=\pi$ is observed.  It has been proposed that the
correlations on the away-side may be the result of a mach-cone induced
by the fast moving away-side parton as it traverses the
medium~\cite{mach}. Other proposals suggest the off-axis peak is due
to Cerenkov radiation from the away-side jet~\cite{cerenkov} or
deflection of jets in the medium~\cite{deflect}.

In this talk, I argue that the ridge and away-side ``cone'' may be a
manifestation of $v_n$ fluctuations; the distinction between two- or
few-particle correlations and $v_n$ fluctuations being that $v_n$
fluctuations are a property of the bulk-medium. The ridge and cone
might not be related to a single hard or semi-hard scatterred parton,
but instead might be due to acoustics in the expansion of the
quark-gluon plasma initiated from a highly textured initial energy
density; one having hot-spots and regions of over- and
under-densities. This hypothesis will need to be further developed to
test for consistency with the full set of heavy-ion data including the
$\Delta\eta$, charge, and particle type dependence of correlations.

\section{A Toy Monte-Carlo}

A hydrodynamic expansion leads to strong space-momentum correlations:
spatial density gradients and interactions lead to boosts which are
largest in the direction of the largest gradient. Information about
the spatial gradient is therefore transferred into momentum space via
the boost driven by the pressure. A system that starts from lumpy
initial conditions and then undergoes a hydrodynamic
expansion~\cite{hydro} should contain non-trivial two-particle
correlations in the final momentum space~\cite{Voloshin:2003ud}. The
conversion of the initial eccentricity $\varepsilon = \frac{\langle y
  - x\rangle}{\langle y+x\rangle}$ into $v_2$ has been discussed
extensively in the literature; see Ref.~\cite{psv} and references
therein. The manifestation of smaller scale spatial fluctuations into
$v_n$ fluctuations, however, is a novel topic. I show results from a
toy model to illustrate the connection of $v_n$ fluctuations and
azimuthal correlations to fluctuations in the initial overlap
geometry.

The toy model assumes that particles in an event are generated from a
finite number of boosted sources. These boosted sources could be
"hot-spots" left over from high-density regions in the initial overlap
area. The boost velocity depends on the radial and azimuthal position
of the source. The number of sources is taken to be $N_{part}$ from a
Monte Carlo Glauber model~\cite{mcg}. The coordinates of the sources
are also taken from the Monte Carlo Glauber model. A blast-wave source
function~\cite{bw} is sampled enough times for each source point so
that the total multiplicity produced in our toy model matches the
$\sqrt{s_{_{NN}}}=$ 200 GeV Au+Au data~\cite{Adams:2004bi}. We
simulate particles within $|\eta|<1$. The details of the model will be
published in an upcoming publication.

Applying a pure Hubble boost to sources distributed according to the
distribution of participants in a Glauber model will lead to negative
$v_2$ values. Typically the blast-wave model is used to model particle
emission from a freeze-out surface which can be highly deformed from
the initial geometry. The deformation is assumed to arise from
rescattering and flow. This toy model is constructed to study how
correlations in the initial conditions can be reflected in
correlations and fluctuations in the produced particles. Since in our
model, the freeze-out geometry is fixed by the initial distributions,
we must apply a boost with a large azimuthal asymmetry in order to
obtain reasonable $v_2$ values. We do not attempt to model the
expansion or to evolve the hot spots with a dynamic model. Such a
calculation has been carried out elsewhere~\cite{Takahashi:2009na}.

\begin{figure}[htb]
\centering\mbox{
\vspace{-15pt}
\includegraphics[width=0.6\textwidth]{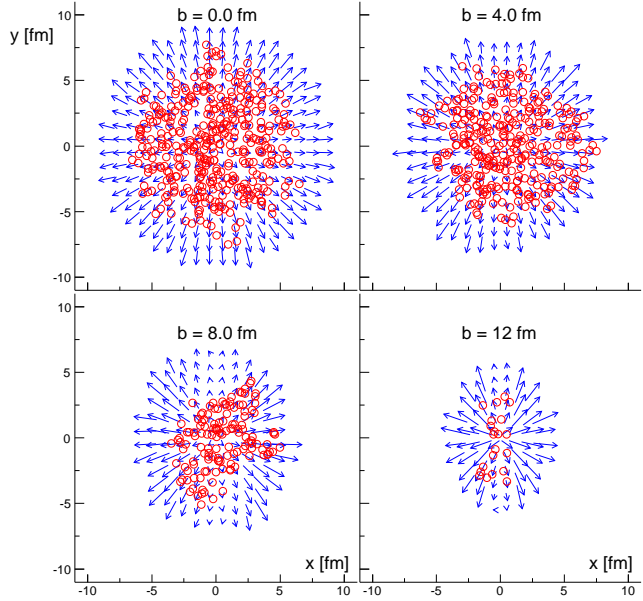}}
\vspace{-15pt}
\caption[]{ A representation of the space-momentum correlations
  induced in this toy Monte Carlo. The arrows represent the mean
  transverse momentum vectors of particles emitted from $x$, $y$
  points in the collision overlap region. The four panels show four
  different impact parameters: $b =$ 0, 4, 8, and 12 fm. }
\label{boost}
\end{figure}

Figure~\ref{boost} shows how the boost employed in our model
influences the momentum of the produced particles. The arrows in the
figure show the average transverse momentum vector for particles
emitted from sources at $x$ and $y$. This model clearly leads to the
desired space momentum correlations since particles emitted from
positive $x$ have a preference for having momentum in the same
direction. The boost also leads to the desired $v_2$ with most arrows
pointing preferentially in the direction of the shortest axis as
anticipated for a pressure driven expansion. The red points show the
locations of participants in one event. The points have been shifted
so that $\langle x\rangle$ and $\langle y\rangle$ are centered at
0,0. This shift has a major effect on $v_1$ fluctuations, reducing the
value of $\langle v_1^2\rangle$ significantly. The uneven distribution
of participants in a single event indicates that, in the case that
space-momentum correlations develop in the expansion of the system,
the produced particles should possess non-trivial correlations.

\section{Model Results}

\begin{figure}[htb]
\centering\mbox{
%\vspace{-15pt}
\includegraphics[width=0.5\textwidth]{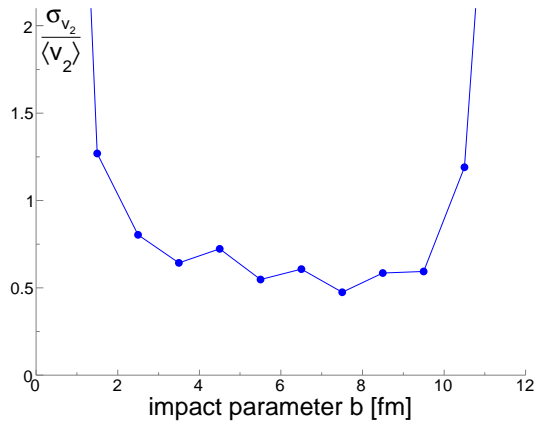}}
%\vspace{-15pt}
\caption[]{ The ratio of $v_2$ fluctuations over the mean as a
  function of impact parameter. }
\label{sigv}
\end{figure}

The model parameters have been tuned so that the integrated $v_2$
matches RHIC data~\cite{Adams:2004bi}. The odd $\langle v_n\rangle$
terms from the model are zero at mid-rapidity but we will see that the
odd $\langle v_n^2\rangle$ terms can be finite.  First, in
fig.~\ref{sigv}, I show the variance of $v_2$ relative to the mean
$v_2$. This ratio has been used in comparisons of data to model
calculations for eccentricity fluctuations~\cite{fluc}. The data in
this figure should be compared with data using $v_{2}$ with respect to
the reaction plane and not the participant plane.  These model results
are similar to what has been observed in preliminary data. We do not
compare to published STAR data~\cite{Adams:2004bi} because that data
was calculated by integrating $v_{2}\{4\}(p_T)$ weighting by the
spectra which is not equivalent to what is shown in this figure. The
ratio is very large in central and peripheral collisions where the
denominator approaches zero. In the intermediate region, the ratio has
a minimum of about 0.5 at $b=7$~fm.

\subsection{Azimuthal Correlations}

\begin{figure}[htb]
\vspace{0.1cm}
\begin{minipage}[b]{0.5\linewidth}
\centering
\includegraphics[width=1.\textwidth]{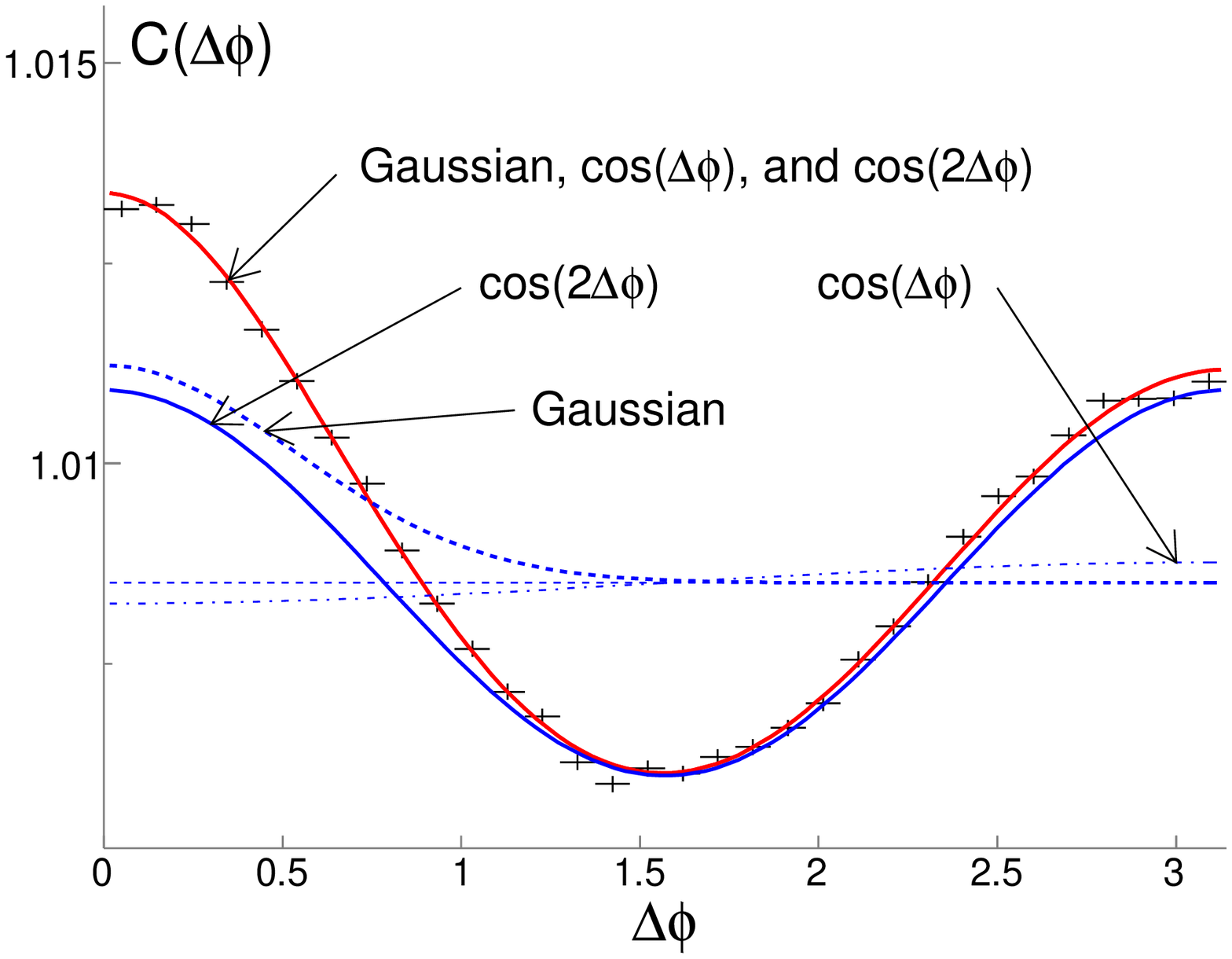}
\end{minipage}
%\hspace{0.01cm}
\begin{minipage}[b]{0.5\linewidth}
\centering
\includegraphics[width=1.\textwidth]{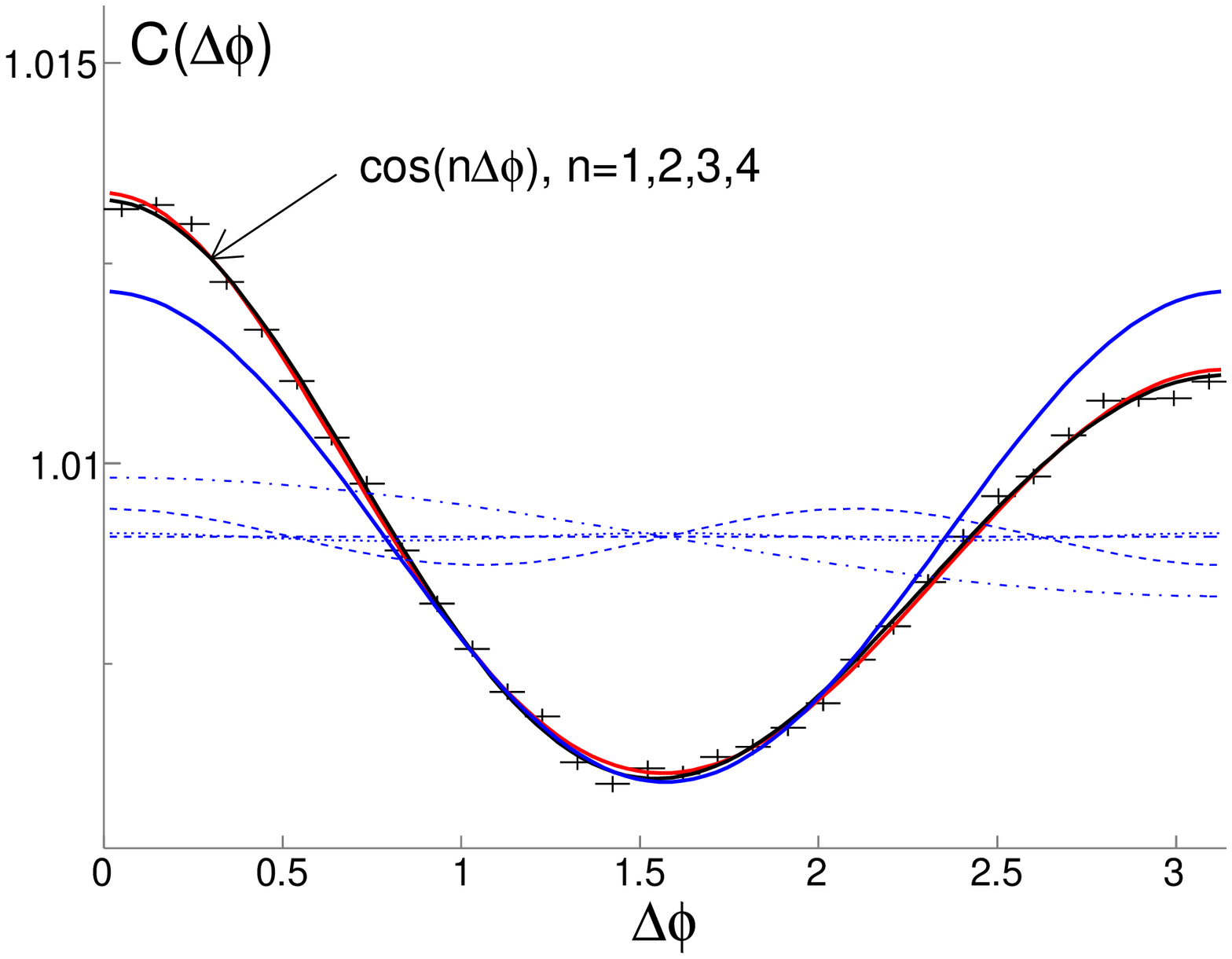}
\end{minipage}
\caption[]{ Both panels: the correlation function for all particles
  produced in the Monte-Carlo with b=6 fm. Left panel: Monte-Carlo
  data fit with a five parameter fitting function including a
  Gaussian, an offset, and two cosine terms. Right panel: data fit
  with a five parameter fitting function including only an offset and
  cosine terms.}
\label{dphi}
\end{figure}

Two-particle azimuthal correlations for all unique pairs of particles
from our model with $b=6$ fm are shown in Fig.~\ref{dphi}. The
correlation function $C(\Delta\phi) = \rho/\rho_{ref}$ where $\rho$ is
the pair density and the reference $\rho_{ref}$ is
$\frac{N_{bins}}{N_{events}}\frac{2}{M(M-1)}$ ($M$ is the average
multiplicity). In the left panel, the correlations are fit to a Gaussian
peak centered at $\Delta\phi=0$, a $\cos(\Delta\phi)$, a
$\cos(2\Delta\phi)$ term, and a constant offset (Gaussian fit). In the
bottom panel, the same model results are fit with
$A(1+\Sigma_n2a_n\cos(n\Delta\phi))$ for $n=1,2,3$ and 4 (Cosine
fit). $a_n$ denotes the fit parameters which should be equal to
$\langle v_n^2\rangle + \delta_n$. Both functions have the same number
of fit parameters and give equally good descriptions of the
correlation function. Fits similar to the Gaussian fit have been used
to parametrize data in terms of a near-side mini-jet peak plus two
cosine terms for momentum conservation and for elliptic
flow~\cite{onset}.

Neither of the fits returns the correct value for $\langle
v_2^2\rangle$. The Cosine fit comes closest but overestimates the
value by 4.6\% while the Gaussian fit underestimates the value by
18\%. The Gaussian fit underestimates $\langle v_2^2\rangle$ because
part of $v_1^2$ and $v_2^2$ along with all of the higher order $v_n^2$
terms are subsumed into the Gaussian peak which over-estimates the
contribution from non-flow in our model. We use the true values of
$v_n^2$ in our model to extract the true nonflow that arise from
correlations between particles emitted from the same participant
pair. The true nonflow peak (not shown) is well described by an offset
and a narrow Gaussian with a width of 0.54 radians. The $n=2$
component of the the true nonflow accounts exactly for the differences
between the known $\langle v_2^2\rangle=1.451e-03$ and the parameter
$a_2=1.517e-03$ in the Cosine fit. This exercise demonstrates that the
parameters extracted from a fit to a correlation function are not
easily related to $\langle v_n^2\rangle$. The results of this
simulation contradict the conclusion in Ref.~\cite{Trainor:2009gj}
that $\langle v_2^2\rangle$ is typically over-estimated in di-hadron
analyses and that a Gaussian type fit can be used to extract $\langle
v_2^2\rangle$ with minimal systematic error. Even when that fit uses
the $\Delta\eta$ dependence, it requires the assumption that
$v_n(\eta_1)v_n(\eta_2)$ has no covariance:
$E(v_n(\eta)v_n(\eta+\Delta\eta))=E(v_n(\eta)v_n(\eta))$. Such an
assumption can mimic a Gaussian dependence for the nonflow in
$\Delta\eta$. In this simulation, the Gaussian fit underestimates
$\langle v_2^2\rangle$ and overestimates the nonflow correlations. It
is best for model builders to compare
their models to measured correlation functions and experiments
should publish unmanipulated correlation functions.

\subsection{$v_n$ Fluctuations}

\begin{figure}[htb]
\centering\mbox{
%\vspace{-15pt}
\includegraphics[width=0.6\textwidth]{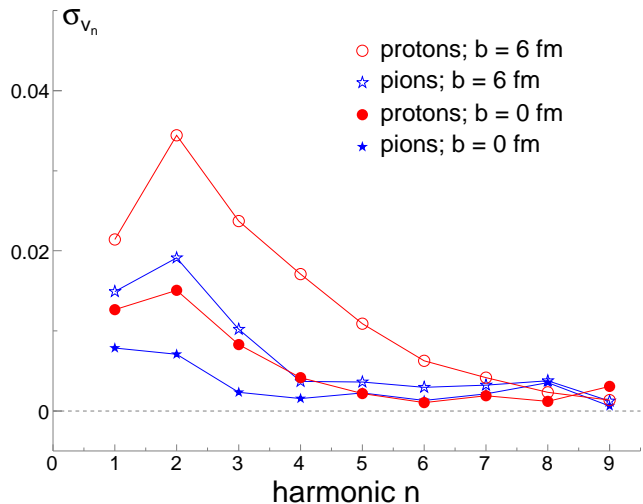}}
%\vspace{-15pt}
\caption[]{ The variance or $RMS$ of the $v_n$ distributions for $n$ from
  1 to 9. Results are shown for protons and pions at impact parameter
  $b=0$ and 6 fm. }
\label{vnrms}
\end{figure}

Fig.~\ref{vnrms} shows $v_n$ fluctuations as a function of $n$ where
$\sigma_{v_n}=\sqrt{\langle v_n^2\rangle - \langle
  v_n\rangle^2}$. Results are shown for pions and protons produced
from events with impact parameters of 0 and 6 fm. As anticipated, both
the even and the odd terms of $\sigma_{v_n}$ are non-zero. Except for
$n = 1$, the values of $\sigma_{v_n}$ generally drop with $n$. The
$n=1$ term is suppressed because we applied a shift to the $x$ and $y$
of the participants in our model so that $\langle x\rangle=0$ and
$\langle y\rangle=0$ (re-centering the events). This significantly
reduces $\sigma_{v_1}$ by constraining the geometric fluctuations, so
that for most cases $\sigma_{v_2}>\sigma_{v_1}$. The $\sigma_{v_n}$
values are larger for protons than for pions. This comes about in our
model because geometry fluctuations drive the $v_n$ fluctuations and
heavier particles carry more information about the geometry
(space-momentum correlations are larger when the ratio of the particle
mass to freeze out temperature $m/T$ is larger).

The non-zero values of $\sigma_{v_n}$ for odd terms and the falling
trend of $\sigma_{v_n}$ with $n$ implies that $v_3$ fluctuations
should not be neglected. In this model, the $v_3$ fluctuations are
particularly prominent for protons and at higher momentum where the
space-momentum correlations are strongest. $v_3$ fluctuations provide
a ready explanation for the double-hump structure on the away-side of
di-hadron correlations. $v_n$ fluctuations also provide a ready
explanation for why correlations on the near- and away-side of
di-hadron correlations both increase rapidly together with
centrality~\cite{onset}; they are actually the same correlation
function arising from the bulk but artificially divided into
components.

\subsection{Relationship to Strangeness and Conclusion}

At this conference, several speakers have discussed how correlations
and fluctuations affect $p_T$ spectra. Fits using Tsallis statistics
have proved successful for fitting the spectra of strange and heavy
flavor hadrons~\cite{tsallis} (the focus of this conference). The same
lumpy initial energy density that leads to the $v_n$ fluctuations
presented here, can also be the source of the temperature fluctuations
implied by the Tsallis fits. If this is true, one should be able to
construct a model that describes all these phenomena at once. That
model of bulk QCD matter should transition smoothly to a pQCD picture
when enough energy is focused within a small enough region.

In this talk, I've shown results from a toy Monte-Carlo model that
illustrate the potential importance of $v_n$ fluctuations in
understanding the initial conditions and expansion of heavy-ion
collisions. I showed that lumpy initial conditions coupled with
space-momentum correlations developed in an expansion phase, leads to
correlations and fluctuations similar to those seen in RHIC data.

{\bf Acknowledgements}
I am grateful to the conference organizers for the invitation to speak
at SQM in Buzios. I am particularly grateful to Eduardo Fraga, Jun
Takahashi, and Takeshi Kodama.

\end{document}